\documentclass[%
 reprint,
nofootinbib,
 amsmath,amssymb,
 aps,
]{revtex4-2}
\usepackage{graphicx}
\usepackage{dcolumn} 
\usepackage{bm}      
\usepackage{enumerate}
\usepackage{amsthm}
\usepackage[T1]{fontenc}
\usepackage[utf8]{inputenc}
\usepackage{lmodern}
\usepackage{makecell}
\usepackage{dblfloatfix}
\usepackage{float}  
\usepackage{ulem}
\usepackage{caption}

\usepackage{array}      
\usepackage{tabularx}
\usepackage{booktabs}

\usepackage[dvipsnames]{xcolor}
\usepackage{hyperref}
\hypersetup{
    colorlinks,
    linkcolor={red!50!black},
    citecolor={blue!50!black},
    urlcolor={blue!80!black}
}

\newcommand{\Nu}{\mathrm{N\mkern-2mu u}}
\newcommand{\Ra}{\mathrm{R\mkern-2mu a}}

\makeatletter
\newcommand*{\rom}[1]{\expandafter\@slowromancap\romannumeral #1@}
\makeatother

\renewcommand{\thefootnote}{\fnsymbol{footnote}}

\begin{document}

\preprint{APS/123-QED}

\title{Freezing lakes as analogue models of \texorpdfstring{\\}{} \texorpdfstring{$\Lambda$}{}CDM cosmology and beyond}

\author{Lorens F. Niehof$^{*,\dagger}$}
\affiliation{%
 Department of Mathematics and Computer Science \\
 Eindhoven University of Technology, 5600~MB Eindhoven, The Netherlands
}
\affiliation{
Department of Applied Physics and Science Education,\\
Eindhoven University of Technology, 5612~AP Eindhoven, The Netherlands}

\author{Ananya Venkatasubramanian$^{*,\ddagger}$}

\author{Federico Toschi$^{\S}$}

\affiliation{
Department of Applied Physics and Science Education,\\
Eindhoven University of Technology, 5612~AP Eindhoven, The Netherlands}

\author{Stefano Liberati$^{\P}$}

\affiliation{SISSA, International School for Advanced Studies, via Bonomea 265, 34136 Trieste, Italy}
\affiliation{INFN, Sezione di Trieste, via Valerio 2, 34127 Trieste, Italy}
\affiliation{IFPU, Institute for Fundamental Physics of the Universe, via Beirut 2, 34014 Trieste, Italy}

\date{\today}

\begin{abstract}
\noindent We extend previous conduction-based analogies between ice growth in a lake and cosmological expansion by incorporating buoyancy-driven heat transport. Reformulating the Stefan problem with both conductive and convective fluxes yields an evolution equation for the ice thickness $s(t)$ that is structurally analogous to the Friedmann equations for the cosmological scale factor $a(t)$. Beyond reproducing radiation-, matter-, and curvature-like behaviors, we introduce a reduced description of convection in which the vertically integrated heat flux reaching the moving ice–water interface is modeled as a power-law function of the instantaneous liquid-layer thickness, generating two additional effective contributions. The first is a constant term, directly analogous to a cosmological constant, arising from the persistence of buoyancy-driven transport under geometric confinement. The second is a $s^{-1}$ contribution originating from the coupling between the moving ice boundary and the convective boundary layer. This term reflects the specific reduced flux-height Ansatz adopted, rather than a universal physical prediction. When expressed in Friedmann-like cosmological form, this term entails a fluid with negative energy density and equation-of-state parameter $w=-2/3$. In cosmology this term may be an effective one associated to a network of domain walls made of exotic energy/matter, but it might also arise from an energy exchange between cosmological components. Overall, the results should be interpreted as a structural analogy between evolution equations, showing how nonlinear transport mechanisms in a classical moving-boundary problem can reproduce the hierarchy of scaling terms familiar from cosmology within a reduced and analytically tractable framework.

\end{abstract}

\maketitle
\footnotetext[1]{These authors contributed equally to this work.}
\footnotetext[2]{\textcolor{blue}{l.f.niehof@student.tue.nl}}
\footnotetext[3]{\textcolor{blue}{a.venkatasubramanian@student.tue.nl}}
\footnotetext[4]{\textcolor{blue}{f.toschi@tue.nl}}
\footnotetext[5]{\textcolor{blue}{liberati@sissa.it}}

\setcounter{footnote}{0}
\renewcommand{\thefootnote}{\arabic{footnote}}

\section{Introduction}\label{sec:introduction} 
The use of analogue systems has become an established tool in theoretical physics for probing regimes of cosmology and gravitation that are otherwise inaccessible. {Examples include analogue black holes and cosmological geometries engineered in Bose–Einstein condensates and optical fibers, which have been used to mimic phenomena such as Hawking radiation from quantum field theory in curved spacetime. Classical systems, including gravity waves in shallow-water flumes, have likewise been employed to probe key aspects of matter dynamics in curved-spacetime analogues \cite{Barcelo2003Analogue, Barcelo2011AnalogueGravity, Volovik2003Universe}.} Such approaches provide conceptual bridges across disciplines and offer both pedagogical and experimental platforms for exploring the behavior of fundamental equations in tangible settings. A broad survey of such analogies between cosmology and classical physical systems can be found in~\cite{FaraoniBookCosmicAnalogies}. Within this tradition, recent work has highlighted geophysical processes, such as glacier flow and lake freezing, as natural analogues of cosmological expansion~\cite{Vollmer2019Freezing, Faraoni2020Analogy}.

In cosmology, the Friedmann--Lemaître--Robertson--Walker (FLRW) equations encapsulate how different energy components (radiation, matter, curvature, and dark energy) govern the time evolution of the scale factor $a(t)$ \cite{Weinberg1972Gravitation, Peebles1993Principles, Mukhanov2005Foundations, Dodelson2020Modern, Planck2018Parameters}. Each component contributes a distinct scaling law, producing successive epochs in cosmic history. Analogously, the growth of an ice layer on a lake is described by the Stefan problem, where heat transport occurs by conduction through the ice and convection in the underlying water \cite{Stefan1889Theory, Carslaw1959Conduction, Alexiades1993Mathematical}. These processes likewise generate contributions with characteristic dependencies on the ice thickness $s(t)$, inviting a structural comparison with FLRW dynamics.

Vollmer \cite{Vollmer2019Freezing} introduced a simplified Stefan model for the freezing of lakes as a pedagogical example of conduction-driven ice growth. 
The structural analogy between this conduction-only model and the radiation-dominated Friedmann equation was subsequently developed by Faraoni \cite{Faraoni2020Analogy}, who showed explicitly that the Stefan equation can be written in Friedmann-like form. Yet conduction alone cannot mimic the full richness of cosmic history, in particular the transition from decelerating to accelerating expansion. Extending the analogy therefore requires incorporating convection, the dominant mode of heat transport in natural lakes and a fundamentally nonlinear process.



In this work we pursue two extensions of this idea. First, we analyze a constant-flux convection model for the freezing-lake problem, which alongside the conductive term yields effective matter- and curvature-like contributions. Second, we develop a reduced description of buoyancy-driven transport beneath the ice by introducing a flux--height Ansatz in which the vertically integrated convective heat flux reaching the moving interface is modeled as a power-law function of the instantaneous liquid-layer thickness.

Convective heat-transfer analogies with cosmology have been explored previously in other thermal systems \cite{Faraoni2020Convective}, where Newtonian cooling and generalized power-law heat fluxes were shown to produce Friedmann-like evolution equations. The present work differs in that it embeds these ideas within a Stefan moving-boundary problem describing lake freezing, allowing the interplay between conduction, convection, and interface motion to generate additional effective terms in the analogue Friedmann equation.

By reformulating ice-growth dynamics into Friedmann-like form, we reveal a close structural correspondence between lake-freezing and cosmic expansion. The analogy not only offers a transparent pedagogical tool but also highlights how identifiable mechanisms in fluid transport can generate effective terms reminiscent of those postulated in cosmology.

Our study can  also serve as a guide for the choice of specific analogue thermal fluid systems that can lead to different cosmological models.

The remainder of this paper is organized as follows. Section II reviews the cosmological background, Section III summarizes the conduction-only Stefan model, Sections IV and V present the constant-convection and flux--height ice growth extensions, and Section VI discusses implications, limitations, and future directions.

\section{Cosmological Expansion}\label{sec:sec2} 

A standard working assumption is that the large-scale dynamics of the Universe is well described by general relativity and that, on sufficiently large scales, the Copernican principle holds, i.e.~spatial homogeneity and isotropy. These symmetries imply a natural foliation of spacetime by spacelike hypersurfaces of constant curvature (i.e.\ maximally symmetric spatial slices). 

They are embodied in the Friedmann--Lemaître--Robertson--Walker (FLRW) metric~\cite{Friedmann1922Curvature, Robertson1935Kinematics, Walker1936Milne}, which describes the geometry in terms of a (time dependent) scale factor $a(t)$ applied to such constant curvature hypersurfaces
\begin{equation}
\mathrm{d} s^{2}
= -\mathrm{d}t^{2}
+ a^{2}(t)\left[
\frac{\mathrm{d}r^{2}}{1-k r^{2}}
+ r^{2}\mathrm{d}\Omega^{2}
\right],
\end{equation}
where $k=0,\pm1$ determines the spatial curvature (flat, closed or open) and $\mathrm{d}\Omega^{2}=\mathrm{d}\theta^{2}+\sin^{2}\theta\,\mathrm{d}\phi^{2}$ is the line element on the unit 2-sphere.
Inserting this metric into the Einstein field equations yields the Friedmann equations, the central relations that dictate how the scale factor evolves in time~\cite{Weinberg1972Gravitation,Peebles1993Principles}.

The first Friedmann equation,
\begin{align}
\label{eq:Friedmann}
    \left(\frac{\dot{a}}{a}\right)^2 = \frac{8\pi G}{3} \rho - \frac{k c^2}{a^2} + \frac{\Lambda c^2}{3},
\end{align}
relates the Hubble rate $H(t) = {\dot{a}}/{a}$ to the total energy density $\rho$, the spatial curvature $k$, and the cosmological constant $\Lambda$. Each constituent of the cosmic energy budget contributes a term with distinct scaling behavior: radiation scales as $a^{-4}$, matter as $a^{-3}$, curvature as $a^{-2}$ and dark energy remains constant with respect to $a$ \cite{Mukhanov2005Foundations,Dodelson2020Modern}. 

The second Friedmann equation,
\begin{align}
    \left(\frac{\ddot{a}}{a}\right) = -\frac{4\pi G}{3} \left(\rho + \frac{3 p}{c^2}\right) + \frac{\Lambda c^2}{3},
\end{align}
encodes the acceleration or deceleration of expansion, depending on the effective equation of state $p = w \rho c^2$ of the dominant component and its strength with respect to the cosmological constant. In particular,
\begin{equation}
    \rho(a) = \rho_{0} \left( \frac{a}{a_0} \right)^{-3(1+w)} \,,
\end{equation}
where normally the reference scale factor $a_0$ is taken to be the scale factor at present and set equal to one, and the reference density $\rho_0$ is taken to be the presently observed one for each cosmological component. The different scaling of the cosmological components with the universe expansion leads to a layered structure of cosmic history, radiation domination at early times, followed by matter domination and eventually dark-energy-driven acceleration~\cite{Planck2018Parameters}. 

In the lake-freezing problem, conduction and convection contribute terms to the ice growth rate equation that similarly scale with inverse powers of the ice thickness $s(t)$. For example, conduction generates a $s^{-4}$-like contribution, formally analogous to radiation domination, while convective processes introduce terms with the same scaling as matter or curvature contributions, as we demonstrate later. In the Flux--height (FH) ice growth model of Section V, additional terms arise, including constant and damping contributions that resemble, respectively, a cosmological constant and exotic fluids absent in the standard $\Lambda$CDM framework.

By casting the growth of lake ice into differential equations structurally homologous to the Friedmann equations, we highlight how familiar terrestrial processes can mirror the mathematical form of cosmological dynamics. This analogy does not imply a physical equivalence between the systems but rather provides an interpretive bridge: the sequential dominance of conduction and convection in ice growth parallels the successive cosmic epochs driven by different energy components. As such, the freezing lake becomes a tractable model system in which to explore, visualize, and reinterpret the behavior of expanding universe solutions.

Beyond these epochal transitions, one can also speculate about more fundamental phase changes in the fabric of spacetime itself. The phase transition from liquid water to solid ice offers an evocative parallel for contemplating phase changes in the early universe. Just as the microscopic structure of water reorganizes abruptly at the freezing point, one can imagine that the fabric of spacetime itself may have undergone a transition between distinct ``phases'' governed by different underlying degrees of freedom. In cosmology, such transitions are hypothesized in scenarios like the Hagedorn phase of string gas cosmology, where the universe shifts from a dense string-dominated regime to a lower-energy state \cite{Hagedorn1965Interactions, Brandenberger1989Superstrings, Brandenberger1989StringGas}, or in models where spacetime emerges from more fundamental pre-geometric constituents \cite{Padmanabhan2008Emergent, Verlinde2011Origin, Volovik2003Universe, Barcelo2011AnalogueGravity,Gielen:2013kla, Gielen:2013naa}. Viewing the freezing of a lake through this lens highlights how macroscopic behavior (such as the sudden appearance of rigidity) can emerge from collective microscopic dynamics, offering an accessible analogy for how radically different physical laws might govern successive phases of the universe’s evolution.

\section{Conduction-only Stefan Model (Vollmer)}
Before turning to convective extensions, it is instructive to recall the conduction-only Stefan model introduced by Vollmer~\cite{Vollmer2019Freezing}. The cosmological interpretation of this model was later developed by Faraoni~\cite{Faraoni2020Analogy}, who showed that the Stefan equation governing purely conductive ice growth can be recast in Friedmann-like form corresponding to a radiation-dominated universe.  This model establishes the baseline for the lake-freezing analogy used throughout the present work.


The Stefan problem describes the advance of a phase boundary in terms of an energy balance between latent heat and conductive flux. In one spatial dimension, the latent heat released as the ice thickens by an amount $dz$ balances the conductive flux through the ice layer
\begin{equation}
\rho L_f \,\frac{dz}{dt} = \frac{\lambda_i (T_w - T_a)}{z},
\end{equation}
where $\rho$ is the density of ice, $L_f$ the latent heat of fusion, $\lambda_i$ the thermal conductivity of ice, $T_w$ the temperature at the ice-water interface (taken as $0\,^\circ\mathrm{C}$), $T_a$ the ambient air temperature, and $z(t)$ the ice thickness.

Introducing the scaled thickness
\begin{equation}
s \equiv \frac{z}{\lambda_i},
\end{equation}
the growth law becomes
\begin{equation}
\dot{s} = \frac{\alpha}{s}, \qquad \mbox{with}\quad
\alpha \equiv \frac{T_w - T_a}{\rho L_f \lambda_i}.
\end{equation}
Squaring and dividing by $s^2$ yields
\begin{equation}
\left(\frac{\dot{s}}{s}\right)^2 = \frac{\alpha^2}{s^4}. \label{eq:vollmer_friedmann}
\end{equation}

Equation~\eqref{eq:vollmer_friedmann} is structurally identical to the first Friedmann equation in a radiation-dominated universe,
\begin{equation}
\left(\frac{\dot{a}}{a}\right)^2 \propto \frac{1}{a^4},
\end{equation}
with $s(t)$ playing the role of the cosmological scale factor $a(t)$. The solution,
\begin{equation}
s(t) \propto t^{1/2},
\end{equation}
exactly mirrors the scaling law $a(t)\propto t^{1/2}$ of the radiation era in standard cosmology.

This correspondence highlights why the lake-freezing problem provides such a compelling analogue: a simple conduction-driven Stefan problem recovers the same functional dependence as the radiation-dominated universe. At the same time, the analogy is limited: conduction alone cannot generate matter- or dark-energy-like terms, nor can it reproduce the transition from decelerated to accelerated expansion. These limitations motivate the convective extensions developed in the following sections.

\section{Constant-convective Model}\label{sec:model1}
A natural first extension of the purely conduction-driven Stefan problem \cite{Stefan1889Theory,Carslaw1959Conduction,Alexiades1993Mathematical,Vollmer2019Freezing,Faraoni2020Analogy} is to include a convective heat flux from the liquid water to the ice interface. Convective heat-transfer laws have previously been used to construct cosmological analogies in other thermal systems \cite{Faraoni2020Convective}; here we apply this idea specifically to the freezing-lake Stefan problem with a moving phase boundary. This idealization amounts to assuming a fixed water-side heat transfer coefficient $h_w$, independent of ice thickness and time. While physically simplistic, this model captures the essential interplay between conduction through the ice and convection in the water, and yields an evolution law for the ice thickness structurally identical to the Friedmann equation with radiation-, matter-, and curvature-like terms.

\begin{figure}[ht]
    \centering
    \includegraphics[width=0.48\textwidth]{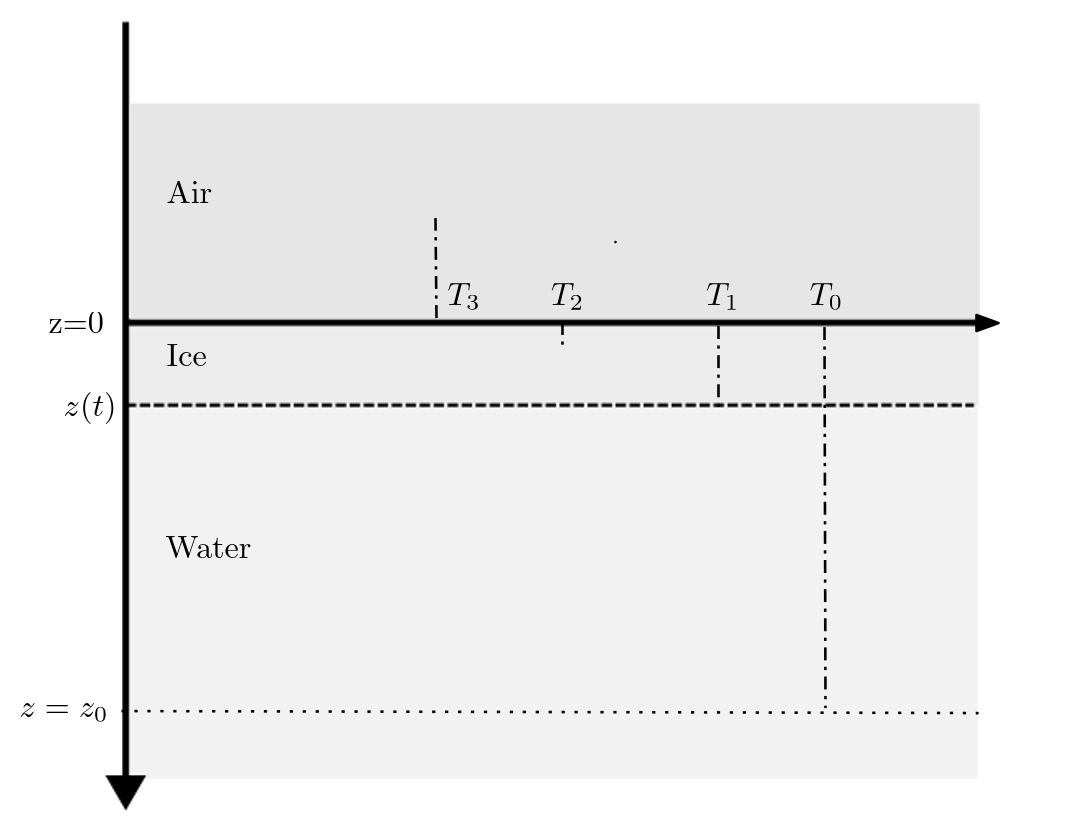}
    \caption{Schematic representation of the temperatures. Here, $T_0$, $T_1$, $T_2$ and $T_3$ denote the bulk water, ice–water interface, ice-air interface and ambient air temperatures, respectively. }
    \label{fig:schematic}
\end{figure}

We consider a one-dimensional vertical geometry with ice thickness $z(t)$. Thermal energy is transferred through conduction within the ice, by convection and radiation at the air-ice interface, and by convective heat input from the water-ice interface. We adopt a quasi-steady-state approximation for the temperature profile, assuming that the conductive and air-side (convective plus radiative) heat fluxes balance at all times. Under this condition, the latent heat released at the ice-water boundary equals the total upward heat loss:
\begin{align}
    q_{\text{cond}} &= \frac{\lambda_i (T_1 - T_2)}{z} \equiv h_{\text{cond}} (T_1 - T_2), \\
    q_{\text{conv+rad}} &= h (T_2 - T_3), \\
    q_{\text{conv+rad}} &= q_{\text{cond}} \equiv q_{\text{ice} \to \text{air}},
\end{align}
where $\lambda_i$ is the thermal conductivity of ice, $h$ is the effective air-side heat transfer coefficient, and $T_0$, $T_1$, and $T_3$ denote the bulk water, ice–water interface, and ambient air temperatures, respectively. Figure \ref{fig:schematic} depicts a rough schematic of the one dimensional vertical air-ice-water model with temperatures. 

This balance allows the elimination of the intermediate temperature $T_2$, yielding the composite heat flux through the ice and air layers as
\begin{align}
    q_{\text{ice}\to\text{air}} = \frac{(T_1 - T_3)}{\tfrac{1}{h} + \tfrac{z}{\lambda_i}}. 
\end{align}
On the water side, the convective heat flux is modeled as
\begin{align}
    q_{\text{water}\to\text{ice}} = h_w (T_0 - T_1),
\end{align}
where the coefficient $h_w$ represents the heat transfer from the liquid to the ice interface.

The latent heat balance at the phase boundary is therefore
\begin{equation}\label{eq:latent_balance}
    L_f \rho \frac{dz}{dt} = q_{\text{ice}\to\text{air}} + q_{\text{water}\to\text{ice}},
\end{equation}
where $L_f$ and $\rho$ are the latent heat and density of ice, respectively. Substituting the expressions for the fluxes gives
\begin{equation}
    \frac{dz}{dt} = \frac{1}{\rho L_f}\!\left[\frac{T_1 - T_3}{\tfrac{1}{h} + \tfrac{z}{\lambda_i}} + h_w (T_0 - T_1)\right].
\end{equation}

Introducing the scaled variable
\begin{equation}
    s \equiv \frac{z}{\lambda_i} + \frac{1}{h},
\end{equation}
and defining constants
\begin{equation}
    \alpha \equiv \frac{T_1 - T_3}{\rho L_f \lambda_i}, \qquad
    \beta \equiv \frac{h_w (T_0 - T_1)}{\rho L_f \lambda_i},
\end{equation}
the evolution equation becomes
\begin{equation}
    \dot{s} = \frac{\alpha}{s} + \beta. \label{eq:simple_first}
\end{equation}
This relation expresses the balance between conductive and convective heat transfer in terms of an effective growth variable $s(t)$ that combines geometric and thermal resistance effects.

Squaring and dividing by $s^2$ yields a Friedmann-like equation,
\begin{align}\label{eq:simple_friedmann}
    \left(\frac{\dot{s}}{s}\right)^2 = \frac{\alpha^2}{s^4} + \frac{2\alpha\beta}{s^3} - \frac{(-\beta^2)}{s^2}.
\end{align}
The three contributions scale as $s^{-4}$, $s^{-3}$, and $s^{-2}$, directly analogous to radiation, matter, and spatial curvature in cosmology.  
The corresponding asymptotic regimes are
\begin{equation}
s(t)\propto
\begin{cases}
t^{1/2}, & \text{radiation-like regime}, \\
t^{2/3}, & \text{matter-like regime}, \\
t, & \text{curvature-like regime}.
\end{cases}
\end{equation}

Differentiating Equation~\eqref{eq:simple_first} with respect to time gives 
\begin{equation}
    \ddot{s} = -\frac{\alpha}{s^2} \dot{s} = -\frac{\alpha}{s^2} \left(\frac{\alpha}{s} + \beta\right) = -\frac{\alpha^2}{s^3} - \frac{\alpha \beta}{s^2},
\end{equation}
which then gives the analogue of the cosmological acceleration equation\footnote{Note that in the cosmological context the acceleration equation can be obtained in the same way from the first Friedmann equation after using the continuity equation to simplify.}:
\begin{equation}
    \frac{\ddot{s}}{s} = -\frac{\alpha^2}{s^4} - \frac{\alpha\beta}{s^3}. \label{eq:simple_accel}
\end{equation}
Notably, there is no $s^{-2}$ term, mirroring the structure of the second Friedmann equation. 

\begin{figure}[ht]
    \centering
    \includegraphics[width=0.48\textwidth]{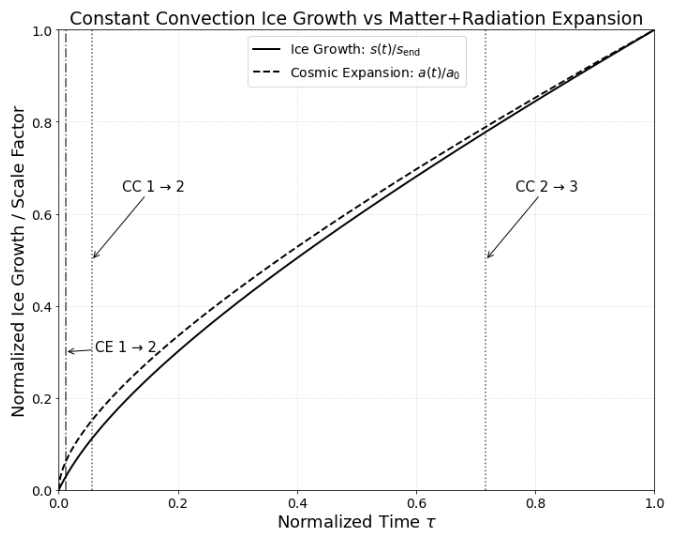}
    \caption{Normalized ice thickness $s(t)/s_{\mathrm{end}}$ in the \textbf{constant-convective (CC)} ice-growth model (solid curve) compared with the normalized scale factor $a(t)/a_0$ in a matter–radiation \textbf{cosmological expansion (CE)} model with zero cosmological constant ($\Lambda = 0$), both plotted as functions of normalized time $\tau$. The analogy reproduces the successive radiation-, matter-, and curvature-like regimes but lacks late-time acceleration. The identified regime transitions for the CC model are: CC$\,1 \rightarrow 2$: $\tau \approx 0.0555$, $s \approx 2.0010$; and CC$\,2 \rightarrow 3$: $\tau \approx 0.7168$, $s \approx 8.0016$. For the CE model, the single transition is CE$\,1 \rightarrow 2$: $\tau \approx 0.0112$, $a \approx 0.0601$. The numbers (1, 2, 3) denote the dominant terms in each regime. Reference constants are provided in Appendix~\ref{app:sim}.}
    \label{fig:ccplot}
\end{figure}

While remarkable, it is worth noticing that this model has some limitations. First of all, it is easy to see that the analogy works only for the case of negative curvature spacetimes as one has to identify $k=-\beta^2$. 
Another issue is that the coefficients of the different analogue cosmological components in  \eqref{eq:simple_friedmann} are evidently not independent, which would correspond to precise relations among the present day densities of the cosmological fluids. Finally, in this simple model, only decelerating dynamics are possible unless $\alpha$ changes sign, corresponding to melting rather than freezing. Thus, while pedagogically transparent and mathematically elegant, the constant-convective model cannot reproduce late-time acceleration (at least not without introducing a time dependence in $\alpha$). In this sense, the constant-convective model is equivalent to a pure CDM universe without cosmological constant $\Lambda$ or any form of dark energy (a fluid for which $\rho+3p<0$). The evolution of the ice thickness in the model is compared with that of the scale factor in a simple CDM universe in Figure~\ref{fig:ccplot}.

\section{Flux--Height Ice Growth Model}\label{sec:model2}
The assumption of a constant water-side heat flux is unrealistic for natural lakes, where buoyancy-driven convection depends on the thickness of the liquid layer beneath the ice. Rather than attempting to resolve the full convective flow, we adopt a reduced description in which the net convective heat flux reaching the ice--water interface is expressed as a function of the instantaneous height of the convecting layer.

Our point of departure is the phenomenological Ansatz
\begin{equation}\label{eq:flux_height_ansatz}
    q_{\mathrm{conv}} \propto H^{\mu}
\end{equation}
where $H \equiv z_0 - z$ is the thickness of the buoyant water layer and $\mu$ is an effective exponent. This form should be understood as an Ansatz for the vertically integrated heat flux entering the Stefan condition, not as a statement about the microscopic structure of the convective flow.
The physical motivation of Equation \eqref{eq:flux_height_ansatz} is threefold.

First, in Stefan problems with convection, the interface position evolves on timescales much longer than the convective turnover time. This separation of timescales allows the convective heat flux to be treated as quasi-steady and be parametrized by the instantaneous geometry of the convecting layer.

Second, the quantity entering the ice-growth equation is the total heat flux delivered to a moving boundary, which depends not only on local transport efficiency but also on the available vertical extent over which buoyancy-driven motion can develop.

Third, when turbulent transport, boundary-layer structure, and geometric confinement are combined into a reduced description, it is natural for the resulting flux to obey an effective power-law dependence on $H$, even if the underlying flow obeys more complex scaling relations.

This Ansatz is consistent with, but more general than, the classical Rayleigh--B\'enard phenomenology. In standard convection theory, the convective flux is often written as
\begin{equation}
    q_{\mathrm{conv}} = \frac{\Nu \lambda_w (T_0 - T_1)}{H},
\end{equation}
with $\Nu$ the Nusselt number and $\lambda_w$ the thermal conductivity of water. If $\Nu \propto Ra^{\gamma}$ and $\Ra \propto H^{3}$, then even classical Rayleigh--B\'enard scaling implies\footnote{In fact, taking the typical Rayleigh--B\'enard scaling $\gamma \sim \tfrac{1}{3}$ reduces to the same Friedmann-like equation as the constant convective model of Section \ref{sec:model1}. }
\begin{equation}
    q_{\mathrm{conv}} \propto H^{3\gamma - 1}.
\end{equation}
In this sense, the exponent $\mu$ in Equation \eqref{eq:flux_height_ansatz} should be interpreted as an effective flux--height exponent that subsumes both transport efficiency and geometric effects, rather than as a bare Nusselt-Rayleigh exponent\footnote{Rigorous bounds on $\Nu$--$\Ra$ scaling \cite{DoeringConstantin1996} apply to statistically stationary Navier-Stokes convection in fixed domains. The present system involves a time-dependent convective layer height and a moving phase boundary, and in addition we think of a generalized fluid with a generic $Nu$ vs. $Ra$ relation for which no comparable bound on the effective flux--height relation would apply.}.

Adopting Equation \eqref{eq:flux_height_ansatz}, we write the convective contribution to the Stefan condition as
\begin{equation}
    q_{\text{water}\to\text{ice}} = B (z_0 - z)^{\mu},
\end{equation}
where $B$ collects material parameters and temperature differences. Introducing again the scaled thickness
\begin{equation}
    s \equiv \frac{z}{\lambda_i} + \frac{1}{h},
\end{equation}
the evolution equation for the interface becomes
\begin{equation}\label{eq:rb_first}
    \dot{s} = \frac{\alpha}{s} + \beta\,(C-\lambda_i s)^{\mu},
\end{equation}
with $C \equiv z_0 - \frac{1}{h}$ and $\alpha, \beta$ constants defined as in Section \ref{sec:model1}.

The exponent $\mu$ controls which effective contributions appear in the analogue Friedmann equation obtained by squaring Equation~\eqref{eq:rb_first} and dividing by $s^2$. For generic values of $\mu$, the resulting expression contains a proliferation of terms with different (and generally non-integer) powers of $s$, not all of which admit a clear or independent cosmological interpretation. In this sense, $\mu$ plays a structural role in selecting which scaling contributions can be meaningfully identified within the analogue framework. However, selecting
\begin{equation}
    \mu = 1
\end{equation}
should be understood as structurally minimal rather than empirically exact: it corresponds to the lowest-order flux--height dependence that yields a closed and interpretable hierarchy of scaling terms. For this choice, the squared evolution equation produces exactly five contributions with scalings $s^{-4}$, $s^{-3}$, $s^{-2}$, $s^{-1}$, and $s^{0}$ respectively, each of which can be placed in one-to-one correspondence with standard or extended components of a Friedmann-like expansion history.

With this choice, the resulting analogue Friedmann equation reads
\begin{equation}\label{eq:rb_friedmann}
    \left(\frac{\dot{s}}{s}\right)^2 =
    \frac{\alpha^2}{s^4}
    + \frac{2\alpha\beta C}{s^3}
    - \frac{(2\alpha\beta \lambda_i-\beta^2 C^2)}{s^2}
    - \frac{2\beta^2 C\lambda_i}{s}
    + \beta^2\lambda_i^2.
\end{equation}

With this interpretation, the role of the exponent $\mu$ is structural rather than empirical: the chosen value of $\mu$ selects the minimal flux--height dependence required to reproduce the hierarchy of effective cosmological components, without implying any violation of classical Rayleigh--B\'enard theory.

\begin{figure}[ht]
    \centering
    \includegraphics[width=0.48\textwidth]{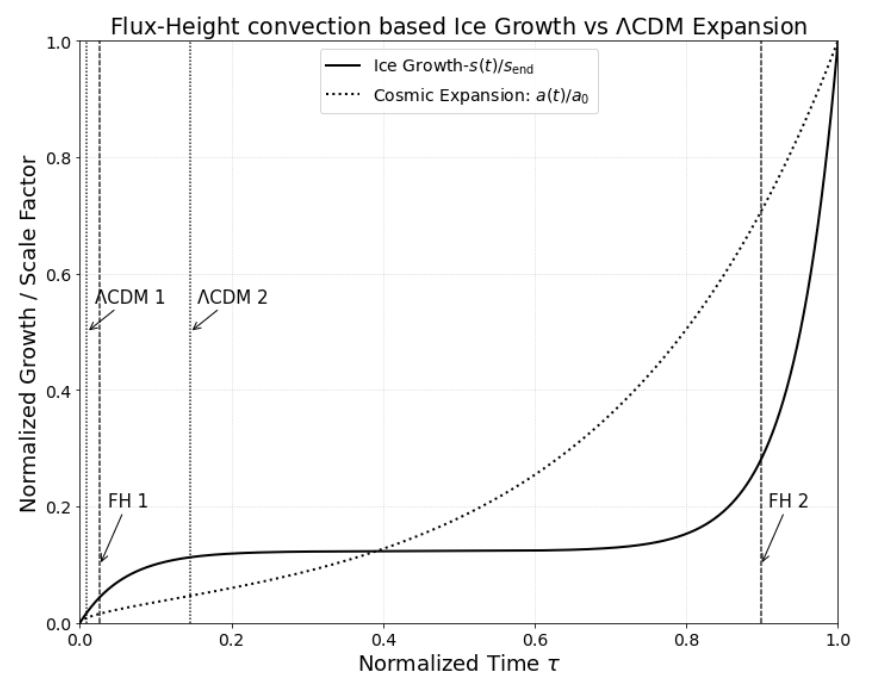}
    \caption{Normalized ice thickness $s(t)/s_{\mathrm{end}}$ in the \textbf{Flux--Height (FH) ice growth} convection model (solid curve) compared with the normalized scale factor $a(t)/a_0$ in the standard \textbf{$\Lambda$CDM} cosmology (dashed curve), both plotted as functions of normalized time $\tau$. The FH model reproduces radiation-, matter-, curvature-, and $\Lambda$-like regimes, while also exhibiting an additional $-s^{-1}$ contribution with no analogue in $\Lambda$CDM, leading to richer dynamical behavior. The observed transition values are as follows: for FH ice growth, $\tau \approx 0.026$, $s \approx 2.291$ (term $3 \rightarrow 4$) and $\tau \approx 0.899$, $s \approx 9.153$ (term $4 \rightarrow 5$); for $\Lambda$CDM expansion, $\tau \approx 0.009$, $a \approx 0.175$ (radiation $\rightarrow$ matter) and $\tau \approx 0.146$, $a \approx 0.931$ (matter $\rightarrow$ dark energy). Reference constants are provided in Appendix~\ref{app:sim}.}
    \label{fig:rbplot}
\end{figure}

\subsection{Interpretations of the \texorpdfstring{$-s^{-1}$}{s⁻¹} contribution}
\label{subsec:sminusone}
Before discussing possible cosmological interpretations, it is important to stress that the appearance of the $s^{-1}$ term is model-dependent and reflects the specific reduced flux--height Ansatz adopted here. Its significance lies not in its literal physical realization, but in illustrating how coupling between a moving boundary and transport efficiency can generate formally exotic contributions when recast in Friedmann-like form.

We just discussed how in the analogue Friedmann--like expression derived from the flux--height (FH) ice growth convection model (Equation \eqref{eq:rb_friedmann}), a distinctive $s^{-1}$ contribution arises, behaving as a perfect fluid with $w=-2/3$ and negative energy density.

The interpretation of this term is far from trivial. If one neglects the sign of its energy density it would be naturally associated in cosmology to a static network of domain walls~\cite{VilenkinShellard1994,Jaeckel2021}. Domain--wall matter, however, ordinarily carries positive energy density and therefore contributes $+a^{-1}$ to $H^{2}$. Our analogue therefore corresponds to a domain--wall--like component but made of exotic negative energy density.

This correspondence is noteworthy because domain walls are a well-studied form of topological defect in cosmology. They arise naturally through the spontaneous breaking of discrete symmetries in scalar-field potentials with multiple degenerate minima, producing two-dimensional surfaces that separate distinct vacuum domains ~\cite{Zeldovich1974Cosmological, Kibble1976Topology, Vilenkin1981Gravitational, Vilenkin1985Cosmic, Battye1999Evolution}. The resulting defect network, behaving as a fluid with tension-like stress and $w=-2/3$, redshifts more slowly than matter but (obviously) more rapidly than a cosmological constant. Although a dominant domain-wall energy density is excluded by cosmic microwave background observations 
\cite{Press1989Cosmology, Larsson1997CMB, Sousa2015Domain, Sousa2020Cosmological}, 
recent analyses incorporating Planck 2018 temperature and polarization data together with BICEP/Keck B-mode measurements have further strengthened these constraints and provided updated bounds on the tension of stable domain-wall networks 
\cite{Caloni2026CMBWalls}. 
Subdominant or transient wall components have nevertheless been discussed in models of early-universe phase transitions, dark energy, and modified-gravity scenarios 
\cite{Bucher1999CMB, Avelino2005Domain, Carter2005Dynamics, Sousa2016Dynamics}. In this context, the appearance of a $w=-2/3$ scaling term in the FH model captures, at an effective-dynamics level, the characteristic behavior of such cosmological defects.


An alternative but equally suggestive interpretation is that this term is not associated to a single exotic cosmological fluid component, but it is rather the by product of the interaction between cosmological components. This idea is not as far fetched as it might seem, given that for example interacting fluid models between dark matter and dark energy have been advanced for resolving the so-called cosmological constant problem (see e.g.~\cite{Barcelo:2006cs, vanderWesthuizen:2023hcl} and references therein). Here we shall limit ourself at providing, in Appendix~\ref{app:interacting}, an explicit toy model, reproducing this negative density $a^{-1}$ contribution, arising as an extra effective fluid due to the interaction of dark matter and dark energy.

The interactive origin of the new exotic term seems to be corroborated also by the fact that in the FH model, the $s^{-1}$ term has a specific dynamical origin rooted in how the same moving ice-water interface constrains conductive and convective heat transport. As the ice thickens, the conductive path length lengthens while the height of the buoyant convective layer decreases. When the Stefan condition is applied to this coupled flux and the interface position is differentiated, the dependence of both contributions on the same variable $s$ generates a mixed term $s^{-1}$. 

Finally, the negative sign is a geometric rather than a mechanistic artifact that physically can be understood as the heat flux reducing the ice thickening due to this coupled geometric behavior. The underlying mechanism of a moving phase boundary whose advancement simultaneously alters conductive resistance and the depth-dependent FH convective flux, is consistent with earlier studies of ice–water interfaces and variable-height convection in \cite{Wettlaufer1997}.

\section{Discussion}\label{sec:discussion} 
The models presented above illustrate how the physical growth of lake ice, governed by conduction through the ice and convection within the underlying water, can be reformulated into dynamical equations structurally analogous to the Friedmann equations of cosmology. This analogy is not a literal physical equivalence but a mathematical and conceptual mapping that provides valuable insight into the role of scaling laws, dominant regimes, and transitions between them.

\subsection{Implications for analogue cosmology}

The constant-flux convection model demonstrates that even a minimal combination of conductive and convective heat transport reproduces radiation- and matter-like eras in the analogue Friedmann equation. A curvature-like contribution emerges from the squared convective term, though only for non-positive curvature. The correspondence between the power-law growth regimes $s\propto t^{1/2}$, $s\propto t^{2/3}$, and $s\propto t$ and the standard cosmological epochs is striking, and highlights the pedagogical value of the analogy: the freezing of a lake offers a tangible illustration of how different effective components control expansion histories.

The flux--height ice growth model extends this picture by allowing the convective heat flux reaching the ice--water interface to depend explicitly on the thickness of the buoyant layer beneath the ice. Rather than invoking a literal Nusselt--Rayleigh scaling, the model adopts a reduced Ansatz in which the vertically integrated convective flux is parametrized as a power law of the instantaneous layer height. This perspective naturally generates a richer set of contributions in the analogue Friedmann equation.

In particular, the emergence of a constant term reflects the persistence of buoyancy-driven heat transport even as the convective layer becomes geometrically constrained. When recast in Friedmann-like form, this residual flux appears as a scale-factor–independent contribution, formally analogous to a cosmological constant. Importantly, this term does not require fine-tuning or exotic assumptions: it arises generically from a flux--height relation in which the convective efficiency saturates with increasing depth.

In addition, the flux--height formulation produces a distinctive $-s^{-1}$ contribution, which has no direct analogue in the standard $\Lambda$CDM framework. This term originates from the geometric coupling between the moving ice–water interface and the shrinking convective domain beneath it. As the interface advances, the conductive path length increases while the available convective height decreases, introducing a mixed contribution to the evolution equation that combines boundary motion and transport efficiency.

Within the cosmological mapping, this term corresponds to an effective component with energy density scaling as $\rho \propto a^{-1}$ and equation-of-state parameter $w=-2/3$. Such a scaling is characteristic of domain-wall networks in cosmology \cite{Vilenkin1985Cosmic,Battye1999Evolution,VilenkinShellard1994,Jaeckel2021}, where two-dimensional topological defects contribute negative pressure and tend to accelerate expansion. In the present analogue system, however, the $s^{-1}$ term enters the Friedmann-like equation with a negative sign, implying an effective negative energy density if the mapping $s\leftrightarrow a$ is maintained.

This sign difference highlights both the reach and the limits of the analogy. In the lake-freezing problem, the negative contribution has a clear physical origin: it reflects boundary-layer feedback and geometric confinement rather than the presence of any exotic negative-mass substance. In this sense, the analogue term encodes a structural correction to the transport law, which translates into a formally exotic component when expressed in cosmological language.

Alternatively, the same mathematical structure can be interpreted in terms of interacting cosmological fluids. As discussed in Appendix \ref{app:interacting}, a component with $\rho\propto a^{-1}$ and negative effective energy density can arise from energy exchange between dark matter and dark energy sectors, depending on the direction of the energy flow. The lake model thus provides a classical realization of a mechanism that, in cosmology, is often introduced phenomenologically.

\subsection{Limitations of the analogy}

The correspondence developed in this work should be understood as a structural analogy between evolution equations, rather than a dynamical or physical equivalence between systems. The aim is not to reproduce the detailed microphysics of cosmology, but to show how familiar transport mechanisms can generate effective scaling terms that formally resemble cosmological components when written in Friedmann-like form. In this sense, the analogy is primarily pedagogical and structural: it highlights how different physical processes can give rise to mathematically similar expansion histories, without implying that the underlying physics is the same.

Despite its elegance, the analogy has important limitations. First, the identification of ice thickness $s(t)$ with the cosmological scale factor $a(t)$ is purely formal; no physical mechanism connects the two systems. This differs from standard analogue-gravity models, where field dynamics on curved backgrounds are simulated, rather than the structure of the Einstein equations themselves. Moreover, cosmological energy densities are associated with fundamental fields and particles, whereas the effective components in the ice-growth problem arise from heat transport across different media.

Second, the lake model is inherently simplified. It assumes one-dimensional geometry, quasi-steady temperature profiles, and neglects lateral flows, salinity effects, wind forcing, and transient diffusion, all of which can be important in real lakes. The flux--height Ansatz, while physically motivated, does not attempt to resolve transitions between different convective regimes or the full complexity of turbulent boundary layers.

In this context, it is important to emphasize that the exponent governing the flux--height relation should be interpreted as an effective phenomenological parameter, not as a fundamental Nusselt--Rayleigh scaling exponent of classical Rayleigh--B\'enard convection. Rigorous bounds and experimental studies constrain the asymptotic $\Nu$--$\Ra$ exponent to values $\lesssim \tfrac{1}{2}$ for statistically steady convection in fixed domains. The present system, by contrast, involves a moving phase boundary and a time-dependent convective layer height, so that the relevant quantity entering the Stefan condition is the vertically integrated heat flux delivered to the interface. When turbulent transport, boundary-layer structure, and geometric confinement are combined into a reduced description, the resulting dependence on layer thickness can differ from canonical $\Nu$--$\Ra$ scaling without violating these bounds.

\subsection{Future directions}
Several extensions suggest themselves. On the fluid-dynamics side, one could explore more detailed flux--height Ansatz derived from direct numerical simulations of buoyancy-driven convection with a moving phase boundary. Two- and three-dimensional models could assess the role of lateral flows, stratification, and plume dynamics, and determine how robust the emergent constant and $s^{-1}$ contributions are to changes in geometry and forcing. Laboratory experiments on controlled freezing in shallow tanks would provide a direct test of the predicted scaling behavior.

Although the present work is primarily formal, it also points to measurable consequences in laboratory systems. A cosmological-constant--like contribution corresponds, in the lake problem, to a persistent offset in the late-time ice-growth rate, while the $s^{-1}$ term predicts systematic deviations from pure conduction scaling as the convective layer thins. Such signatures could be tested experimentally by precise measurements of interface motion and heat flux.

On the cosmological side, the analogy motivates consideration of nonstandard contributions to the Friedmann equations. The appearance of an effective $a^{-1}$ component connects naturally to studies of domain-wall fluids and interacting dark-sector models, while the emergent constant illustrates how complex microphysics can generate effective macroscopic parameters. In this way, analogue systems may provide intuition for how exotic terms can arise without fundamental modifications of gravitational theory.

\subsection{Concluding remarks}
The freezing of a lake is an everyday physical process, yet when reformulated through a flux--height description it reproduces many structural features of cosmological expansion. The constant-flux model captures radiation- and matter-like eras, while the flux--height model inspired by buoyancy-driven convection naturally generates curvature-like, cosmological-constant-like, and exotic contributions. Crucially, these terms can be traced to identifiable transport mechanisms and geometric feedbacks rather than introduced ad hoc.

Together, these results demonstrate how nonlinear transport and moving-boundary problems can act as conceptual laboratories for cosmology. They highlight how effective evolution equations with rich component structure can emerge from reduced descriptions of complex dynamics, underscoring the universality of scaling ideas across fluid dynamics, geophysics, and cosmology.

\begin{acknowledgments}
The authors gratefully acknowledge Stefano Ruffo for stimulating this collaboration.

\end{acknowledgments}

\appendix
\section{Mapping between fluid and cosmological variables}\label{app:mapping}

For clarity, Table~\ref{tab:mapping} summarizes the correspondence between the FLRW cosmological framework and the flux--height ice-growth model developed in this work. The analogy is mathematical rather than physical, but it provides a useful dictionary for interpreting the structure of the emergent terms.

The correspondences in Table~\ref{tab:mapping} follow directly from how each contribution to the heat flux scales with the moving boundary thickness $s(t)$. Heat conduction through the ice produces a contribution scaling as $s^{-4}$, which maps onto a radiation-like component with $\rho \propto a^{-4}$. A convective flux that is independent of the liquid-layer height yields a term scaling as $s^{-3}$, corresponding to a matter-like component with $\rho \propto a^{-3}$. The squared convective contribution then generates a $s^{-2}$ term, formally analogous to the curvature contribution scaling as $a^{-2}$ in cosmology.

Within the adopted flux--height Ansatz, buoyancy-driven convection gives rise to a scale-independent contribution ($\propto s^{0}$, which maps onto a cosmological-constant--like term with $\Lambda \propto a^{0}$. In addition, the geometric coupling between the moving ice--water interface and the shrinking convective layer produces a negative $s^{-1}$ contribution. Under the FLRW mapping, this term corresponds to an exotic component scaling as $-a^{-1}$, reminiscent of effective domain-wall fluids or interacting cosmological sectors.

Taken together, these correspondences make explicit how distinct transport mechanisms and geometric feedbacks in the ice-growth problem translate into effective cosmological components, illustrating how a reduced description of nonlinear heat transport can reproduce the full hierarchy of terms appearing in the Friedmann equations.

\begin{widetext}
\begin{center}

\begin{tabular}{|l|l|}
\hline
\textbf{Fluid system (ice growth with convection)} &
\textbf{Cosmological analogue (FLRW dynamics)} \\
\hline
Ice thickness $s(t)$ (moving phase boundary) &
Scale factor $a(t)$ (cosmic expansion variable) \\[6pt]

Growth rate $\dot{s}/s$ &
Hubble parameter $H(t)=\dot{a}/a$ \\[6pt]

Conductive heat flux ($\propto s^{-4}$) &
Radiation-like component ($\propto a^{-4}$) \\[6pt]

Height-independent convective flux ($\propto s^{-3}$) &
Matter-like component ($\propto a^{-3}$) \\[6pt]

Squared convective contribution ($\propto s^{-2}$) &
Curvature-like term ($\propto a^{-2}$) \\[6pt]

Scale-independent flux contribution ($\propto s^{0}$) &
Cosmological constant–-like term ($\propto a^{0}$, $\Lambda$) \\[6pt]

Boundary-layer feedback ($\propto -s^{-1}$) &
Exotic contribution ($\propto -a^{-1}$) \\[6pt]

Transport-regime transitions &
Cosmological era transitions \\
\hline
\end{tabular}
\captionof{table}{Formal correspondence between the transport contributions in the flux--height ice-growth model and the effective components appearing in the Friedmann equations of cosmology. Each entry lists the dependence on the ice thickness $s$ in the fluid system and the analogous scaling with the cosmological scale factor $a$, illustrating how distinct heat-transport mechanisms map onto effective cosmological energy components.}
\label{tab:mapping}
\end{center}
\end{widetext}

\section{Derivation of \texorpdfstring{$w = -\tfrac{2}{3}$}{}}\label{app:w}
Here we show in detail that the $-1/s$ contribution in Equation \eqref{eq:rb_friedmann} corresponds to a perfect fluid with effective equation-of-state parameter $w=-2/3$.

For a homogeneous, isotropic perfect fluid in an FLRW spacetime the local conservation of energy–momentum,
\begin{align}
    \nabla_{\mu} T^{\mu \nu} = 0,
\end{align}
reduces to the continuity equation
\begin{align}
    \dot\rho + 3H(\rho + p) = 0,
\end{align}
where $\rho(t)$ is the energy density, $p(t)$ the pressure, $H(t) = \frac{\dot{a}}{a}$ is the Hubble parameter and $a(t)$ the scale factor. For a barotropic fluid with equation of state
\begin{align}
    p = w \rho,
\end{align}
this becomes
\begin{align}
    \dot{\rho} + 3H\rho(1+w) = 0.
\end{align}
Using $H = {\dot{a}}/{a}$, we can rewrite the time derivative of $\rho$ as
\begin{align}
    \frac{1}{\rho}\dot{\rho} &= -3 (1 + w) \frac{\dot{a}}{a}.
\end{align}
Integrating both sides gives
\begin{align}
\ln\rho = -3(1+w)\ln a + \text{const},
\end{align}
hence
\begin{align}
\rho \propto a^{-3(1+w)}.
\end{align}

If the term in the analogue Friedmann equation scales as $\rho \propto a^{-1}$ (corresponding to the observed $s^{-1}$ dependence in our mapping), we set the exponent 
\begin{align}
    -3(1+w) = - 1.
\end{align}
Solving for $w$ yields
\begin{align}
    w=-\frac{2}{3}.
\end{align}

The value $w = -\tfrac{2}{3}$ has immediate consequences for the usual energy conditions. Assuming $\rho > 0$:
\begin{itemize}
\item \emph{Null Energy Condition (NEC):} $\rho + p = \rho (1 + w) = \frac{1}{3}\rho \geq 0$ — satisfied.
\item \emph{Weak Energy Condition (WEC):} requires $\rho \geq 0$ and $\rho + p \geq 0$ — both hold, so WEC is satisfied.
\item \emph{Dominant Energy Condition (DEC):} requires $\rho \geq |p|$, i.e. $1 \geq |w| = \frac{2}{3}$ — satisfied.
\item \emph{Strong Energy Condition (SEC):} requires $\rho+p\geq 0$ and $0 \leq \rho + 3p  = \rho (1 + 3w) = - \rho < 0$ — violated.
\end{itemize}

Thus the effective fluid that corresponds to the $s^{-1}$ term has $w = -\frac{2}{3}$: it satisfies the NEC, WEC and DEC for positive $\rho$, but it violates the SEC. For negative $\rho$, however, the effective fluid violates the NEC, WEC, DEC and SEC.

\section{Interacting Fluids and the Emergence of an effective fluid component with \texorpdfstring{$w_{\rm eff}=-2/3,\rho_{\rm eff}<0$}{} }\label{app:interacting}

In this appendix we show that cosmological models with energy exchange between fluids can naturally generate an effective contribution to the Friedmann equation scaling as $\rho \propto a^{-1}$ with negative energy density. For doing so we consider an often conjectured interaction between a cold dark matter component (DM, $w_{\rm DM}=0$) and a dark energy one (DE, constant $w_{\rm DE}< -\tfrac13$). This can be represented by a suitable interaction term $Q$ appearing in the continuity equations of these cosmological components 
\begin{align}
    \dot\rho_{\rm DM}+3H\rho_{\rm DM}&=+Q,  \\
    \dot\rho_{\rm DE}+3H(1+w_{\rm DE})\rho_{\rm DE}&=-Q. 
\end{align}

Let us choose a simple energy exchange rate
\begin{equation}
Q \;=\; q\,H\,a^{-1}, \qquad q\in\mathbb{R},
\end{equation}
and seek solutions of the form (standard pieces $+$ interaction--induced corrections)
\begin{align}
\rho_{\rm DM}(a)&=\rho_{{\rm DM},0}\,a^{-3}+A\,a^{-1}\,,\\
\rho_{\rm DE}(a)&=\rho_{{\rm DE},0}\,a^{-3(1+w_{\rm DE})}+B\,a^{-1}\,.
\end{align}
Using $d/dt=Ha\,d/da$ in the continuity equations gives
\begin{align}
\frac{Q}{H}&=2A\,a^{-1},\\
\frac{Q}{H}&=-(2+3w_{\rm DE})\,B\,a^{-1}\,.
\end{align}
So that
\begin{equation}
A=\frac{q}{2},\quad B=-\frac{q}{\,2+3w_{\rm DE}\,}.
\end{equation}
So, the interaction-induced extra term can be interpreted as an effective fluid with
\begin{align}
\rho_{\rm eff}(a) &\equiv (A+B)a^{-1}
= \frac{3q\,w_{\rm DE}}{2\,(2+3w_{\rm DE})}\,a^{-1},\\
p_{\rm eff}(a) &\equiv w_{\rm DM}(A a^{-1}) + w_{\rm DE}(B a^{-1})
= -\,\frac{q\,w_{\rm DE}}{\,2+3w_{\rm DE}\,}\,a^{-1}.
\end{align}
Therefore, as long as $w_{\rm DE}\neq -2/3$,
\begin{equation}
w_{\rm eff}\equiv \frac{p_{\rm eff}}{\rho_{\rm eff}}=-\frac{2}{3}\,.
\end{equation}
The sign of $\rho_{\rm eff}$ is set by $q\,w_{\rm DE}/(2+3w_{\rm DE})$; choosing the sign of $q$ appropriately yields $\rho_{\rm eff}<0$ if desired.

\section{Simulation Constants}\label{app:sim}
The parameters used in the numerical simulations for the constant-convective (CC) and Flux--Height (FH) ice growth models, as well as for the cosmological expansion analogues, are listed in Table~\ref{tab:constants}. Unless otherwise stated, the values correspond to standard physical constants or representative laboratory estimates for freshwater ice, water, and air under terrestrial conditions.

\begin{widetext}
\begin{center}
\begin{tabular}{|l|l|l|l|}
\hline
\textbf{Symbol} & \textbf{Value} & \textbf{Unit} &
\textbf{Description / Source} \\
\hline
$\rho_{\mathrm{ice}}$ & 917 & kg\,m$^{-3}$ &
Density of ice (freshwater, 0\,$^\circ$C) \cite{harvey2017ice} \\

$L_f$ & $3.34\times10^{5}$ & J\,kg$^{-1}$ &
Latent heat of fusion of ice \cite{lazeroms2018antarcticmelt} \\

$\lambda_i$ & 2.22 & W\,m$^{-1}$\,K$^{-1}$ &
Thermal conductivity of ice \cite{voitkovskii1960ice} \\

$\lambda_w$ & 0.56 & W\,m$^{-1}$\,K$^{-1}$ &
Thermal conductivity of water \cite{nist_ice_thermophysical} \\

$\nu$ & $1.61\times10^{-6}$ & m$^{2}$\,s$^{-1}$ &
Kinematic viscosity of water (0–5\,$^\circ$C) \cite{crc2018} \\

$\kappa$ & $1.40\times10^{-7}$ & m$^{2}$\,s$^{-1}$ &
Thermal diffusivity of water \cite{blumm2007thermophysical} \\

$g$ & 9.81 & m\,s$^{-2}$ &
Gravitational acceleration \cite{Escudier2019} \\

$\mu$ & $1$ & — &
Flux--Height power law exponent \\

$T_0$ & 3.0 & $^\circ$C &
Bulk water temperature (estimate) \\

$T_1$ & 0.0 & $^\circ$C &
Ice–water interface temperature (estimate) \\

$T_3$ & $-20.0$ & $^\circ$C &
Ambient air temperature (estimate) \\

$h$ & 15.0 & W\,m$^{-2}$\,K$^{-1}$ &
Surface convective/radiative heat transfer coefficient
\cite{Vollmer2019Freezing} \\

$h_w$ & 5.0 & W\,m$^{-2}$\,K$^{-1}$ &
Water-side convective heat transfer coefficient (estimate) \\

$z_0$ & 10.0 & m &
Reference depth scale \\
\hline
\end{tabular}
\captionof{table}{Physical constants used in the numerical simulations for the constant-convective (CC) and Flux--Height (FH) ice-growth models, compared with their respective cosmological analogues in Figures~\ref{fig:ccplot} and~\ref{fig:rbplot}.}
\label{tab:constants}
\end{center}
\end{widetext}

\bibliography{main_V1}
\end{document}